\documentclass[preprint,showpacs,showkeywords,preprintnumbers,amsmath,amssymb,superscriptaddress]{revtex4}
\usepackage{graphicx}
\usepackage{dcolumn}
\usepackage{bm}
\begin{document}

\title{Fractional Non-Markovian effect and Newton's 2nd law of motion}

\author{Chun-Yang Wang\footnote{Corresponding author: wchy@mail.bnu.edu.cn}}
\affiliation{Shandong Provincial Key Laboratory of Laser Polarization and Information Technology, College of Physics and Engineering, Qufu Normal University, Qufu 273165, China}
\affiliation{State Key Laboratory of Theoretical Physics, Institute of Theoretical Physics, Chinese Academy of Sciences, Beijing 100190, China}
\author{Shu-Qin Lv}
\affiliation{Shandong Provincial Key Laboratory of Laser Polarization and Information Technology, College of Physics and Engineering, Qufu Normal University, Qufu 273165, China}
\author{Ming Yi\footnote{Co-corresponding author: yiming@wipm.ac.cn}}
\affiliation{Key Laboratory of Magnetic Resonance in Biological Systems, Wuhan Institute of Physics and Mathematics, Chinese Academy of Sciences, Wuhan, China}


\begin{abstract}
We report in this paper a thorough study on the the dynamical mechanics of the fractional Brownian motion systems.
Where several non-trivial properties are revealed such as the abundant non-Markovian effects resulted from the fractional characters of the system.
In general, the dynamics of the fBm system is found to be of a purely Newton's type, despite of the anomalous fractional properties of the system.
\end{abstract}

\keywords{}

\pacs{05.70.Ce, 05.30.-d, 05.40.Ca}

\maketitle

\section{INTRODUCTION}

The Newton's laws of motion are of fundamental importance for the develop of theoretical physics.
Among which the second law is the most remarkable as the cornerstone of modern physics.
Recently, as the the development of research in some singular problems such as anomalous diffusion and the quantum thermodynamics of small systems,
many new understandings on the Newton's laws have been raised.
For example, E. Verlinde proposed in 2011 to interpret the force in Newton’s second law and gravity as entropic forces \cite{enrf}.
Bao and his coauthors defined in 2006 an intermediate dynamics between Newton's and Langevin \cite{intd}.
All these studies have encouraged people to make more and more new explorations on the basic theories of modern physics.

Fractional Brownian motion (fBm) which is an alternative quintessential model for stochastic processes has received wide attention in recent years for its abundance of
singularities \cite{jef1,fbm1,fbm2}.
Enormous efforts have been made on the establishment and development of relevant theories \cite{fbm3,fgn1,chyx}.
However, it is not clear for the dynamical origin of many specific behaviors immersing in the fractionally damping environment.
It is of great meaning to sought whether there lives any speciality that is beyond the Newton's and Langevin.
Despite of the difficulties immersed in the fractional calculus, we find from our recent study that this can be easily achieved from
a briefly mathematical discussion.

Therefore in this letter, we report one of our recent study on the inherent dynamics of a fractional damping system that may induce fBm.
The paper is organized as follows:
in Sec.\ref{sec2}, self-autocorrelations of the particles are obtained by Laplacian solving the fractional Langevin equation of the system.
Where the mean position and mean square displacement of the particles are determined during the exploration of its dynamical details in the reactive trajectory.
In further in Sec.\ref{sec3}, the average acceleration of the diffusing particle is defined and computed for the visualization of the fractional dynamical mechanism immersed in the fBm processes.
Sec.\ref{sec4} serves as the summary of our conclusions.

\section{Fractional Non-Markovian effect}\label{sec2}

Different from the standard Brownian motion (sBm), fBm is generally believed to originate from the fractional Gaussian noise (fGn) \cite{fgn2} whose autocorrelation is characterized by the Hurst exponent $H$.
For which, $H=1/2$ corresponds to the standard Brownian motion, $H<1/2$ and $H>1/2$ denote the sub- or super-diffusive cases, respectively.
Theoretically, no matter fBm or sBm, both of them can be categorized in the Langevin unification of fractional motions \cite{fms1} in the study of anomalous diffusions \cite{ad01,ad02,ad03}.

However, here in this letter, instead of other choices, let us begin from the fractional generalized Langevin equation (FGLE) \cite{fle1,fle2,fle3} which is an alternative approach to fBm
\begin{eqnarray}
m\ddot{x}+m\int^{t}_{0}\eta(t-t')\dot{x}(t')dt'+\partial_{x}U(x)=\xi(t)
\label{fle}
\end{eqnarray}
where $\eta(t)=\eta_{\alpha}t^{-\alpha}/\Gamma(1-\alpha)$ is the frictional kernel with fractional exponent $0<\alpha<1$ and
$\eta_{\alpha}$ the strength constant.
$\xi(t)$ is the fGn and its correlation satisfies the fluctuation-dissipation theorem \cite{fdt1} $\langle\xi(t)\xi(t')\rangle=mk_{B}T\eta_{\alpha}|t-t'|^{-\alpha}$.
Mathematically the FGLE written in the form of Eq.(\ref{fle}) can also be yielded from a generalized system-plus-reservoir Hamiltonian model of harmonic oscillators \cite{hotb}.

The solution of the FGLE, namely also the equation of motion for the diffusing particles, can be easily obtained by a series of Laplacian transformation. For an example, in the particular case of a linear potential $U(x)=-Fx$, we find after some algebra that
\begin{eqnarray}
x(t)=\langle x(t)\rangle_{0}+\frac{1}{m}\int^{t}_{0}H(t-\tau)[\xi(\tau)+F]d\tau,\label{sl1}
\end{eqnarray}
where $\langle x(t)\rangle_{0}=x_{0}+v_{0}H(t)$ denotes the time-dependent mean position of the diffusion particle.
$H(t)=\mathcal{L}^{-1}[s^{2}+s\eta(s)]^{-1}$ is the response function with $\eta(s)=\eta_{\alpha}s^{\alpha-1}$ the Laplacian transform of $\eta(t)$. All these results are reminiscent of the previous ones \cite{intd} except for the differences arise from the fractional friction kernel function.

If the system is close to thermal equilibrium at temperature $T$, the generalized Einstein relation \cite{eirl} relates
the fluctuations of the test particle position in the absence of an external field to its behavior under the influence of a constant force field $F$ (time and space independent).
Among which the mean square displacement (MSD) of the force-free particle can be obtained from setting $t_{1}=t_{2}$ in the autocorrelation
\begin{eqnarray}
\{\langle x(t_{1})x(t_{2})\rangle\}&=&\{x^{2}_{0}\}+\{v^{2}_{0}\}H(t_{1})H(t_{2})+\{x_{0}v_{0}\}\left[H(t_{1})+H(t_{2})\right]+\sigma_{x}^{2}(t),
\end{eqnarray}
where the assumption of $\langle x_{0}\xi(t)\rangle=0$ is used for the meaning of no correlation between the noise and the initial position of the particle.
The symbol $\{\cdots\}$ indicates the average with respect to the initial preparation of the state variables and $\langle\cdots\rangle$ is the average over the noise.
And the variance of $x(t)$ reads
\begin{eqnarray}
\sigma^{2}_{x}(t)=\int^{t}_{0}dt_{1}H(t-t_{1})\int^{t_{1}}_{0}dt_{2}\langle
\xi(t_{1})\xi(t_{2})\rangle H(t-t_{2}).
\end{eqnarray}

From another point of view, the FGLE can be rewritten in an overdamped form by using of the velocity function
\begin{eqnarray}
m\dot{v}+m\int^{t}_{0}\eta(t-t')v(t')dt'+\partial_{x}U(x)=\xi(t),
\label{fle}
\end{eqnarray}
which induces another form of analytical result of the FGLE
\begin{eqnarray}
v(t)=v_{0}h(t)+\frac{1}{m}\int^{t}_{0}h(t-\tau)[\xi(\tau)+F]d\tau,\label{sl2}
\end{eqnarray}
where $h(t)=\mathcal{L}^{-1}[s+\eta(s)]^{-1}$ can be regarded as another form of response function. And succeedingly the two-time velocity correlation function (VCF) of the Brownian particle can be obtained in the force free case as
\begin{eqnarray}
\{\langle v(t_{1})v(t_{2})\rangle\}=\{v^{2}_{0}\}h(t_{1})h(t_{2})+\sigma^{2}_{v}(t).
\end{eqnarray}
where the assumption of $\langle v_{0}\xi(t)\rangle=0$ is used, i.e. there exists no correlation between the thermal noise and the initial velocity.
And
\begin{eqnarray}
\sigma^{2}_{v}(t)=\int^{t}_{0}dt_{1}h(t-t_{1})\int^{t_{1}}_{0}dt_{2}\langle
\xi(t_{1})\xi(t_{2})\rangle h(t-t_{2}),
\end{eqnarray}
is the variance of $v(t)$.
Similar results can also be obtained when the particles are subjected in an external filed. With these correlations, the dynamical properties of the fBm stochastic process can be easily revealed whenever it is force-free or not.

\begin{figure}[ht]
\centering
\includegraphics[scale=0.72]{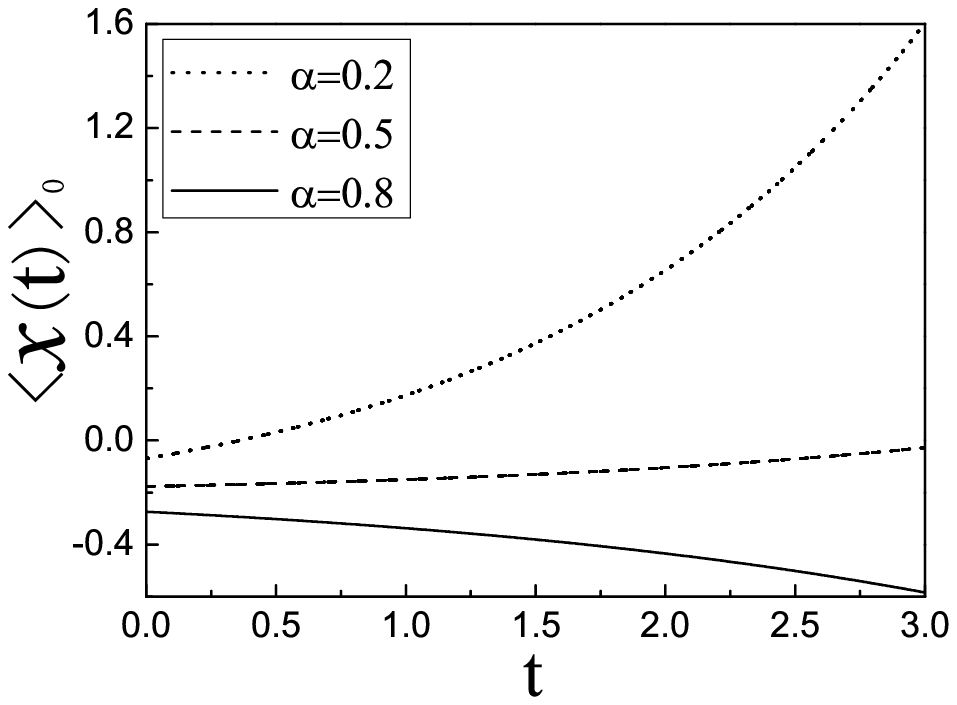}
\includegraphics[scale=0.72]{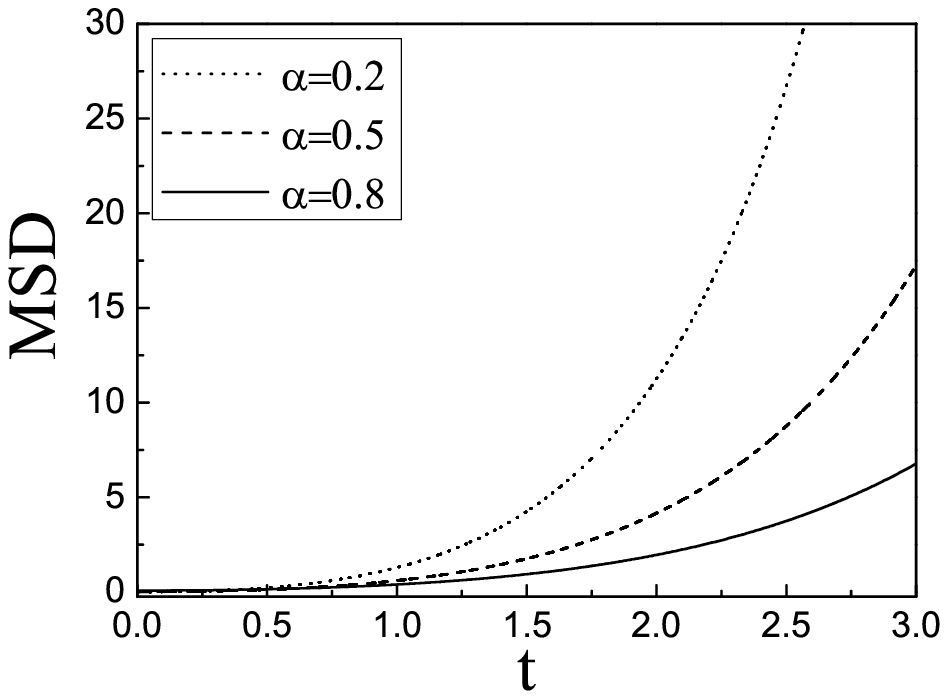}
\caption{Time dependent varying of MSD and MSV at various $\alpha$. Parameters in use are
$\eta_{\alpha}=2.0$, $\omega=k_{B}T=1.0$, $x_{0}=-1.0$ and $v_{0}=2.0$. \label{f1}}
\end{figure}

For an example, for the particle diffusing in a metastable well, the external force suppresses on the particle can be approximated to be $F(x)=-m\omega^{2}x$ in the neighbourhood of the saddle point.
The time-dependent mean position $\langle x(t)\rangle_{0}$ and MSD can then be obtained by a series of computation on the correlations.  As is illustrated in Fig. \ref{f1}, where all the parameters so that dimensionless units such as $k_{B}T=1.0$ can be used for simplicity.
From which we can see that, all the quantities are closely related to the fractional exponent $\alpha$. In particular, $\langle x(t)\rangle_{0}$ is found to change from a positively increasing function to a negatively diverging one as the increasing of $\alpha$.   Which implies an unexpected type of reversely moving in the process of thermodynamic diffusion. This gifts us a thimbleful of insight into the abundant non-Markovian effects immersed in the fBm systems.

\section{Fractional acceleration pursuing and 2nd law}\label{sec3}

But what is mostly concerned here is whether there exists another type of dynamics beyond the Newton's and Langevin as has been discussed in Ref. \cite{intd}.
For this purpose, one should trace back to the two solutions of the FGLE, i.e. Eqs. (\ref{sl1}) and (\ref{sl2}).
Supposing the characteristic equation $s+\eta(s)=0$ possesses a zero root, one can then imply from the residue theorem that
\begin{subequations}
\begin{eqnarray}
h(t)&=&b+\sum_{i}\textrm{Res}[h(s_{i})]\textrm{exp}(s_{i}t),\\
H(t)&=&c+bt+\sum_{i}s_{i}^{-1}\textrm{Res}[h(s_{i})]\textrm{exp}(s_{i}t).
\end{eqnarray}\end{subequations}
Where $s_{i}$ denotes the nonzero roots of the above characteristic equation and $\textrm{Res}[\cdots]$ are the residues of $h(s)=[s+\eta(s)]^{-1}$.
Then we can see here that two relevant coefficients $b$ and $c$ can be defined as well as what has been done previously \cite{intd}.

The average displacement and average velocity of the particle under the external force $F$ can be yielded from
\begin{eqnarray}
\{\langle x(t)\rangle\}&=&\{x_{0}\}+b\left(\{v_{0}\}t+\frac{1}{2}\frac{F}{m}t^{2}\right)+c\left(\{v_{0}\}+\frac{F}{m}t\right)+\frac{F}{m}d,\\
\{\langle v(t)\rangle\}&=&b\left(\{v_{0}\}+\frac{F}{m}t\right)+\frac{F}{m}c,
\end{eqnarray}
in the long time where $d$ is another noise dependent coefficient.
And the acceleration of the fBm particle of mass $m$ can then be obtained as
\begin{eqnarray}
a=\frac{F}{m}b,
\end{eqnarray}
given it is subjected to a linear external field.

The value of $b$ ranges generally from 0 to 1, with $b=1$ corresponding to the purely Newton' mechanics and $b=0$ the trivial Langevin dynamics.
However we noticed that due to the fractional properties of the fBm system, the value of $b$ may should be closely related to the varying of the fractional exponent $\alpha$.
This is what we should pay high attention to.
In order to find the details, we trace back to the definition of all the coefficients.
Yielded from the residue theorem, the determination of coefficient $b$ reads
\begin{eqnarray}
b=\frac{1}{1+\eta'(0)}.\label{b}
\end{eqnarray}
where $\eta'(s)$ is the first order derivative of $\eta(s)$ over $s$.

\begin{figure}[ht]
\centering
\includegraphics[scale=1]{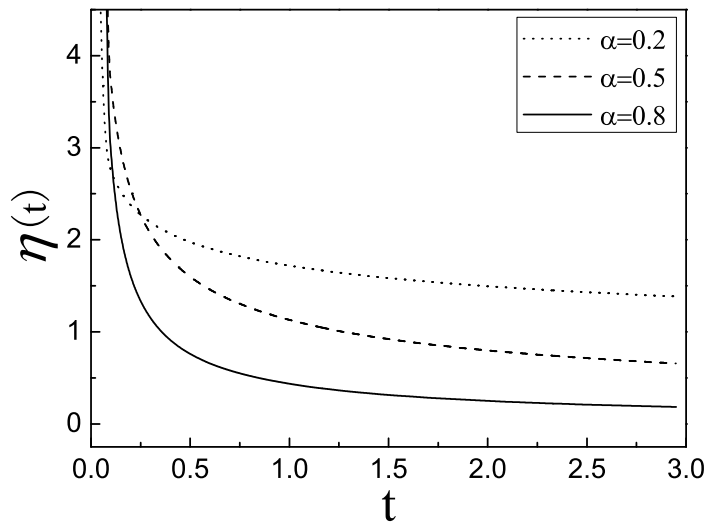}
\caption{Friction kernel $\eta(t)$ plotted as a function of $t$ at various $\alpha$. \label{f2}}
\end{figure}

Due to the closely $s$-dependent of $\eta(s)=\eta_{\alpha}s^{\alpha-1}$, a startling nontrivial result is found that for the fBm system $b=1$ is always met.
This means that the Newton's mechanism is recovered despite of the abundant fractional non-Markovian effects in the fBm system.
Although it is described by the FGLE, the dynamics of the fBm system is not of the Langevin type.
In order to find the intrinsic reasons for this abnormity, the friction kernel function $\eta(t)$ is plotted in Fig. \ref{f2} at various $\alpha$ for a thorough investigation.
From which we can see that, as time goes on $\eta(t)$ evolves quickly to a zero-approximating effective friction.
Thus results in the long time the no dissipation case of the purely Newton's mechanics.
While actually the validity of Eq.(\ref{b}) requires also $\eta(0)=\int_{0}^{\infty}\eta(t)dt=0$, i.e., a vanishing effective friction at zero frequency.

\section{conclusion and prospects}\label{sec4}

In conclusion in this paper we have made a careful study on the dynamical mechanics of the fBm system.
Wherein we have found that the dynamics of the fBm system is of a purely Newton's type, despite of the anomalous fractional properties of the system.
However abundant non-Markovian effects can still be revealed for the fBm system due to its fractional parameters such as the friction kernel function and fractional noise.

Historically, the study concerning on the fractional Brownian motion has been in progress for decades of years.
Since its publication of the original seed by B. Mandelbrot \cite{fbm1}, more than one thousand and seven hundred citations has been indexed.
Among these works, the dynamical property of the fBm system has rarely been considered especially when the system is subjected to an external potential field.
While actually, this is not of great difficulty.
For an example again for the particle diffusing in a metastable well,
the average acceleration of it can be valued from the second order derivative of $\langle x(t)\rangle_{0}=x_{0}+v_{0}H(t)$, i.e., $\langle a(t)\rangle_{0}\propto\frac{d^{2}}{dt^{2}}H(t)$, resulting an average acceleration which is closely related to the fractional exponent $\alpha$ (or in another word, the fractional properties) of the fBm system.
Therefore in prospection, we believe this present study will in some cases stimulate a few in-depth considerations on this project in the near future.

\section * {ACKNOWLEDGEMENTS}

This work was supported by the Shandong Province Science Foundation for Youths under the Grant No.ZR2011AQ016, the Open Project Program of State Key
Laboratory of Theoretical Physics, Institute of Theoretical Physics, Chinese Academy of Sciences, China under the Grant No.Y4KF151CJ1, the National Natural Science Foundation of China under the Grant No.11275259 and 91330113.

\end{document}